\documentclass[]{raa}            % referee version: for submission
\usepackage{graphicx,times}
\usepackage{natbib}

\providecommand{\bm}[1]{\mbox{\boldmath$#1$\unboldmath}}

\begin{document}

   \title{First photometric study of the eclipsing binary PS
   Persei
% $^*$
%\footnotetext{\small $*$ Supported by the National Natural Science Foundation of China.}
}

 \volnopage{ {\bf 2009} Vol.\ {\bf 9} No. {\bf XX}, 000--000}
   \setcounter{page}{1}

   \author{Jinzhao Yuan}
   \institute{Physics and Information Engineering Institute, Shanxi
Normal University, Linfen 041004, China; {\it
yuanjz@sxnu.edu.cn}\\}

\abstract{The CCD photometric observations of the eclipsing binary
PS Persei (PS Per) were obtained on two consecutive days in 2009.
The 2003 version Wilson-Devinney code was used to analyze the
first complete light curves in $V$ and $R$ bands. It is found that
PS Per is a short-period Algol-type binary with the less massive
component accurately filling its inner critical Roche lobe. The
mass ratio of $q=0.518$ and the orbital inclination of
$i=89.^{\circ}86$ are obtained. On the other hand, based on all
available times of primary light minimum including two new ones,
the orbital period has been improved. }

   \authorrunning{Jinzhao Yuan}            %author_head in even pages
   \titlerunning{First photometric study of PS Per }  % title_head in odd pages
   \maketitle

%% The author head (on even pages) and the title head (on odd pages) will be
%% automatically extracted from \author{} and \title{}. Whenever the title is too long,
%% you will be asked to supply a shorter one by inserting either \authorrunning{} or
%% \titlerunning{} before \maketitle. Anyway, you can specify your own heads.
%%
%%
%% Note: In the following text body of your manuscript, please note several differences from
%%       other major journals:
%% (1) \subsection{Please Capitalize the First Letter of Each Notional Word in Subsection Title}
%% (2) Please Capitalize the First Letter of Each Notional Word in all tables' captions

%
%________________________________________________ sections below
%
\section{Introduction}           %% first-level sections will be auto-capitalized
\label{sect:intro}

PS Per ($\alpha_{2000.0}=02^{h}39^{m}33.^{s}3$ and
$\delta_{2000.0}=+45^{\circ}38^{'}05.^{''}5$) was designated by
Kukarkin et al. (1968). But photographic and visual times of light
minimum have been obtained since 1926. Later, Photoelectric and
CCD times of light minimum were published by \v{S}af\'{a}\v{r} \&
Zejda (2000a), \v{S}af\'{a}\v{r} \& Zejda (2000b), Zejda (2002),
Agerer \& H\"{u}bscher (2003), Zejda (2004), Diethelm (2005),
H\"{u}bscher et al. (2006), Zejda et al. (2006), Br\'{a}t et al.
(2009), and Diethelm (2010). But, no complete light curve of the
binary system have been made so far for photometric analysis.

In this paper, the first complete light curves in $V$ and $R$
bands were presented. And the absolute physical parameters as well
as orbital period were determined.

% Authors can give a citation as `Michel et al. 1992'.
% You may also use \cite, \citep and \citet for citation, and use Table~1
% or Figure~1 and so forth. Using \ref and \label for cross-references of
% Tables/Figures is a good way in adjusting/adding/removing text, tables or
% figures.

\section{Observations}
\label{sect:Obs}

New CCD photometric observations of PS Per in $V$ and $R$ bands
were carried out on 2009 November 13 and 14 using the 85-cm
telescope at the Xinglong Station of National Astronomical
Observatory of China (NAOC), equipped with a primary-focus
multicolor CCD photometer. The telescope provides a field of view
of about $16.^{'}5\times16.^{'}5$ at a scale of $0.^{''}96$ per
pixel and a limit magnitude of about 17 mag in $V$ band.

The typical exposure times in $V$ and $R$ bands were 90s and 60s
respectively. The coordinates of the variable, comparison, and
check stars are listed in  Table 1. The data reduction was
performed by using the aperture photometry package
IRAF{\footnote[1]{IRAF is developed by the National Optical
Astronomy Observatories, which are operated by the Association of
Universities for Research in Astronomy, Inc., under contract to
the National Science Foundation.}} (bias subtraction, flat-field
division). Extinction corrections were ignored as the comparison
star is very close to the variable. In total, 445 CCD images in
the $V$ band and 446 images in the $R$ band were obtained. Several
new times of light minimum (see Table 2) are derived from the new
observation by using a parabolic fitting method.

The first complete light curves in $V$ and $R$ bands are obtained,
and displayed in the top panel of Figure 1. The new orbital period
revised in the next section was used to calculate the phase.

\begin{table}[h!!!]
\small \centering
\begin{minipage}[]{120mm}
\caption[]{ Coordinates of PS Per and its Comparison and Check
Stars.}\label{Table 1}
\end{minipage}
\tabcolsep 6mm
\begin{tabular}{lll}
\hline\noalign{\smallskip}
Stars           & $\alpha_{2000}$         &$\delta_{2000}$ \\
\hline\noalign{\smallskip}
PS Per          & $02^{h}39^{m}33.3^{s}$  & $45^\circ38'05.5"$\\
Comparison  & $02^{h}39^{m}24.1^{s}$ & $45^\circ42'22.1"$\\
Check       & $02^{h}39^{m}22.2^{s}$ & $45^\circ43'34.0"$\\
\noalign{\smallskip}\hline
\end{tabular}
\end{table}

\begin{table}[h!!!]
\small \centering
\begin{minipage}[]{120mm}
\caption[]{ New CCD Times of Light Minimum for PS
Per.}\label{Table 2}
\end{minipage}
\tabcolsep 6mm
\begin{tabular}{lllll}
\hline\noalign{\smallskip}
 No.    & J.D. (Hel.) (days) &Error (days)  & Min.  & Filter \\
\hline\noalign{\smallskip}
 1      &2455149.2742        & $\pm0.0004$      & I     & $V$    \\
        &2455149.2752        & $\pm0.0005$      & I     & $R$    \\
 2      &2455149.9768        & $\pm0.0003$      & I     & $V$    \\
        &2455149.9768        & $\pm0.0004$      & I     & $R$    \\
 3      &2455150.3267        & $\pm0.0004$      & II    & $V$    \\
        &2455150.3265        & $\pm0.0004$      & II    & $R$    \\
\hline\noalign{\smallskip}
\end{tabular}
\end{table}

\begin{figure}[h!!!]
\centering
\includegraphics[width=9.0cm,angle=0]{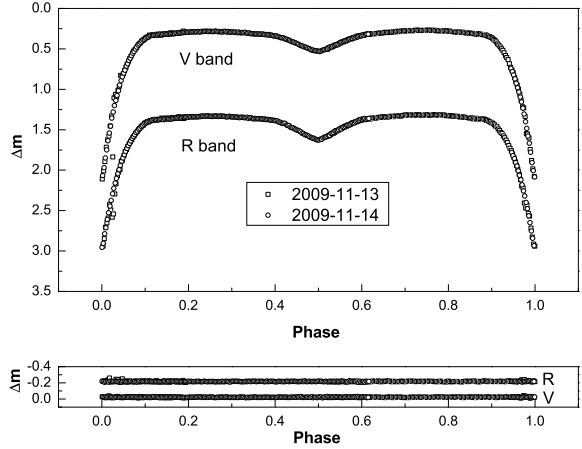}
\begin{minipage}[]{85mm}
\caption{Top panel: the light curves of PS Per in the $V$ and $R$
bands obtained on 2009 November 13 and 14. The points in $R$ band
has been shifted down by 1.0 mag. Bottom panel: the differential
light curves of the comparison star relative to the check star.
The points in $R$ band has been shifted up by 0.2 mag.}
\end{minipage} \label{Fig1}
\end{figure}

\section{Orbital period study}
All available times of primary light minimum seen in literature
were collected and listed in Table 3, which also includes the data
in the database of eclipsing binaries of Kreiner (2004). For my
two-band light minima, a mean time of light minimum is given. The
$O-C$ values of all minimum times were computed with the ephemeris
given by Kreiner et al. (2001):
\begin{equation}
\mathrm{Min}~I= 2424527.2165 + 0^{d}.70217968\times{E},
\end{equation}
and listed in the fifth column of Table 3. The corresponding $O-C$
diagram, Figure 2, shows that the data are distributed around a
straight line. So, a linear ephemeris was used to fit the $O-C$
values. The photographic and visual data show large deviation from
the straight line for their low quality. In the fitting process, a
weight of 10 is used for photoelectric and CCD data, and 1 for
photographic and visual data. The CCD times, 2451841.3121 and
2452213.4662, have the weight of 1 for their large errors. A
least-squares fit to the data gave the following ephemeris:
\begin{equation}
\mathrm{Min}~I= 2424527.2163 + 0^{d}.70217977\times{E},
\end{equation}
The new ephemeris is plotted in Figure 2 with a solid line. The
residuals with respect to Equation (2) are listed in the sixth
column of Table 3.

\begin{table}[h!!!]
\small \centering
\begin{minipage}[]{120mm}
\caption[]{Times of primary light Minimum for PS Per.}\label{Table
3}
\end{minipage}
\tabcolsep 3mm
\begin{tabular}{ccccccc}
\hline\noalign{\smallskip} JD (Hel.) & Method & Error &  E  &
$O-C$ & Residuals  & Ref.\\\hline\noalign{\smallskip}
2424527.250  & p   &       &     0 &  0.03350 &  0.03374 & MVS 2                   \\
2428834.407  & p   &       &  6134 &  0.02034 &  0.02000 & MVS 2                   \\
2430972.531  & p   &       &  9179 &  0.00722 &  0.00660 & MVS 2                   \\
2430991.464  & p   &       &  9206 & -0.01863 & -0.01925 & MVS 2                   \\
2435718.545  & p   &       & 15938 & -0.01124 & -0.01249 & MVS 2                   \\
2435725.534  & p   &       & 15948 & -0.04404 & -0.04529 & MVS 2                   \\
2436114.558  & p   &       & 16502 & -0.02758 & -0.02888 & MVS 2                   \\
2436596.315  & p   &       & 17188 &  0.03416 &  0.03278 & MVS 2                   \\
2436603.330  & p   &       & 17198 &  0.02736 &  0.02598 & MVS 2                   \\
2436850.460  & p   &       & 17550 & -0.00988 & -0.01128 & MVS 2                   \\
2436852.580  & p   &       & 17553 &  0.00358 &  0.00217 & MVS 2                   \\
2436876.438  & p   &       & 17587 & -0.01253 & -0.01393 & MVS 2                   \\
2436895.406  & p   &       & 17614 & -0.00338 & -0.00479 & MVS 2                   \\
2437588.460  & p   &       & 18601 & -0.00073 & -0.00223 & MVS 2                   \\
2437939.528  & p   &       & 19101 & -0.02257 & -0.02412 & MVS 2                   \\
2437944.477  & p   &       & 19108 &  0.01117 &  0.00961 & MVS 2                   \\
2437946.551  & p   &       & 19111 & -0.02136 & -0.02291 & MVS 2                   \\
2437970.442  & p   &       & 19145 & -0.00447 & -0.00602 & MVS 2                   \\
2438321.547  & p   &       & 19645 &  0.01069 &  0.00908 & MVS 2                   \\
2438385.453  & p   &       & 19736 &  0.01834 &  0.01672 & MVS 2                   \\
2447028.565  & v   &       & 32045 &  0.00065 & -0.00211 & BBSAG 85                \\
2447118.446  & v   &       & 32173 &  0.00266 & -0.00011 & BBSAG 86                \\
2447170.406  & v   &       & 32247 &  0.00136 & -0.00142 & BBSAG 87                \\
2447384.566  & v   &       & 32552 & -0.00344 & -0.00625 & BBSAG 89                \\
2447491.294  & v   &       & 32704 & -0.00675 & -0.00957 & BBSAG 90                \\
2447566.432  & v   &       & 32811 & -0.00198 & -0.00481 & BBSAG 91                \\
2447894.348  & v   &       & 33278 & -0.00389 & -0.00677 & BBSAG 94                \\
2448136.597  & v   &       & 33623 & -0.00688 & -0.00979 & BBSAG 96                \\
2448283.362  & v   & .004  & 33832 &  0.00257 & -0.00036 & BBSAG 97                 \\
2448509.458  & v   & .005  & 34154 & -0.00329 & -0.00625 & BBSAG 99                 \\
2448867.573  & v   & .003  & 34664 &  0.00007 & -0.00294 & BBSAG 102                \\
2449202.511  & v   & .004  & 35141 & -0.00163 & -0.00468 & BBSAG 104                 \\
2449546.575  & v   & .006  & 35631 & -0.00568 & -0.00878 & BBSAG 107                 \\
2449653.312  & v   & .003  & 35783 &  0.00001 & -0.00310 & BBSAG 108                  \\
2449945.401  & v   & .005  & 36199 & -0.01774 & -0.02089 & BBSAG 110                  \\
2450713.6127 & cc  & .0014 & 37293 &  0.00939 &  0.00612 & \v{S}af\'{a}\v{r} \& Zejda (2000a)\\
2450721.3309 & cc  & .0008 & 37304 &  0.00362 &  0.00035 & BBSAG 116                         \\
2450839.2981 & cc  & .0021 & 37472 &  0.00463 &  0.00135 & \v{S}af\'{a}\v{r} \& Zejda (2000b)\\
2450841.4035 & cc  & .0021 & 37475 &  0.00349 &  0.00021 & \v{S}af\'{a}\v{r} \& Zejda (2000b)\\
2451077.3366 & cc  & .0006 & 37811 &  0.00422 &  0.00091 & BBSAG 119                     \\
2451088.572  & v   & .003  & 37827 &  0.00474 &  0.00142 & BBSAG 119                     \\
2451515.501  & v   & .008  & 38435 &  0.00850 &  0.00513 & BBSAG 121                     \\
2451810.419  & v   & .005  & 38855 &  0.01103 &  0.00762 & BBSAG 123                     \\
2451841.3121 & cc  & .0058 & 38899 &  0.00823 &  0.00481 & Zejda (2002)                   \\
2451876.4201 & cc  & .0017 & 38949 &  0.00724 &  0.00382 & Zejda (2002)                   \\
2451878.533  & v   & .008  & 38952 &  0.01360 &  0.01018 & BBSAG 124                      \\
2451899.5900 & cc  & .0003 & 38982 &  0.00521 &  0.00179 & Agerer \& H\"{u}bscher (2002)  \\
2452190.295  & v   & .003  & 39396 &  0.00783 &  0.00437 & BBSAG 126                      \\
2452204.3365 & cc  & .0010 & 39416 &  0.00573 &  0.00227 & BBSAG 126                      \\
2452213.4637 & cc  & .0005 & 39429 &  0.00460 &  0.00113 & BBSAG 127                      \\
2452213.4662 & cc  & .0070 & 39429 &  0.00710 &  0.00363 & Zejda (2004)                   \\
2452260.516  & v   & .004  & 39496 &  0.01086 &  0.00739 & BBSAG 127                      \\
2452524.525  & v   & .007  & 39872 &  0.00030 & -0.00320 & Diethelm (2003)                \\
2452531.5507 & pe  & .0002 & 39882 &  0.00420 &  0.00069 & Agerer \& H\"{u}bscher (2003)  \\
2452885.446  & v   & .002  & 40386 &  0.00094 & -0.00261 & Diethelm (2004)                \\
2453302.5428 & cc  & .0010 & 40980 &  0.00301 & -0.00059 & Diethelm (2005)                \\
2453656.4422 & cc  & .0003 & 41484 &  0.00385 &  0.00019 & Zejda et al. (2006)            \\
2453705.5937 & pe  & .0008 & 41554 &  0.00278 & -0.00088 & H\"{u}bscher et al. (2006)     \\
2453988.5713 & cc  & .0001 & 41957 &  0.00197 & -0.00172 & Br\'{a}t et al. (2009)         \\
2454019.4675 & cc  & .0001 & 42001 &  0.00226 & -0.00144 & Br\'{a}t et al. (2009)         \\
2455114.8667 & cc  & .0006 & 43561 &  0.00116 & -0.00268 & Diethelm (2010)                \\
2455149.2747 & cc  & .0003 & 43610 &  0.00236 & -0.00149 & this paper                     \\
2455149.9768 & cc  & .0003 & 43611 &  0.00228 & -0.00157 & this paper                     \\
\hline\noalign{\smallskip}
\end{tabular}
\end{table}

\begin{figure}[h!!!]
\centering
\includegraphics[width=9.0cm,angle=0]{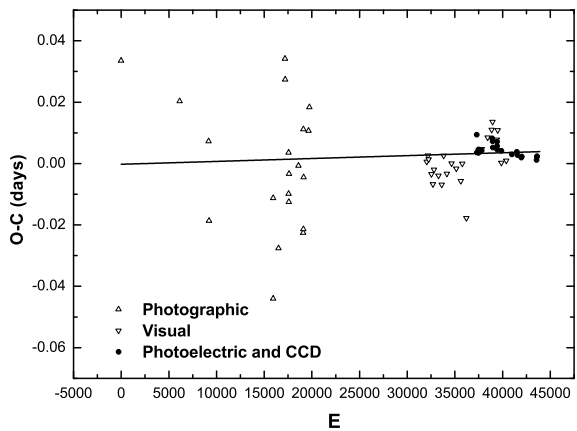}
\begin{minipage}[]{85mm}
\caption{$O-C$ diagrams. The solid line is calculated with the new
ephemeris in Equation (2).}
\end{minipage} \label{Fig2}
\end{figure}

\section{Photometric solutions with the W-D method}
The light curves were analyzed using the 2003 version of the
Wilson-Devinney code (Wilson \& Devinney 1971; Wilson 1979, 1990).
Since the spectral type of PS Per is F5, an effective temperature
of $T_1 = 6750$K is assumed for the primary component. Assuming
the photospheric surface of the binary star is convective,
gravity-darkening coefficients ($g_1=g_2=0.320$) and bolometric
albedo ($A_1=A_2=0.5$) were used. According to the tables of van
Hamme (1993), the limb-darkening coefficients 0.506 for $V$ band
($x_{1V} = 0.506$) and 0.414 for $R$ band ($x_{1R} = 0.414$) were
adopted.

Since no mass ratio has been published in literature, a $q$-search
method was used to determine the mass ratio. Solutions were
carried out for a series of values of the mass ratio
$q=M_{2}/M_{1}$ ($q=0.3$, 0.4, 0.5, 0.6, 0.7, 0.8, 0.9, 1.0).
Considering the light curves of EB type, mode 2 (detached mode) is
assumed. The behavior of the sum of the residuals squared,
$\Sigma$, as a function of mass ratio $q$ is plotted in Figure 3,
showing that $\Sigma$ reaches the minimum value near $q=0.5$.
Therefore, the mass ratio was taken as an adjustable parameter and
given the initial value of $q = 0.5$. After some differential
corrections, The solution converged to mode 5 (semi-detached) and
gave the final mass ratio of $q=0.518$. The derived physical
parameters are listed in Table 4. The theoretical light curves
computed with the parameters are plotted in Figure 4 as a solid
line.

\begin{figure}[h!!!]
\centering
\includegraphics[width=9.0cm,angle=0]{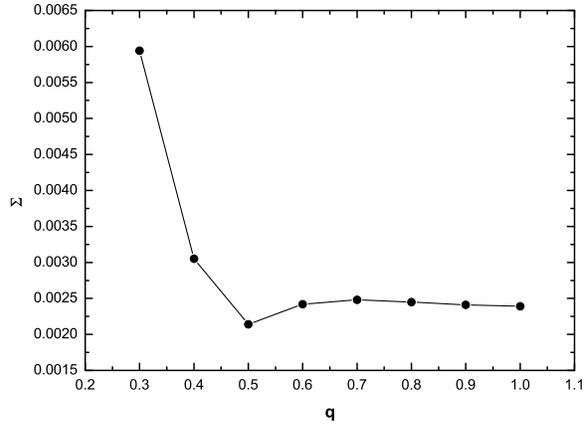}
\begin{minipage}[]{85mm}
\caption{Relation between $\Sigma$ (the sum of the residuals
squared) and $q$ for PS Per.}
\end{minipage} \label{Fig3}
\end{figure}

\begin{table}
\caption{Photometric Solutions for PS Per.}\label{table 4}
\begin{center}
\small
\begin{tabular}{lll}
\hline\noalign{\smallskip}
Parameters                  &        Photometric elements     &     Errors              \\
\hline\noalign{\smallskip}
$g_{1}=g_{2}$               &0.32                             & assumed     \\
$A_{1}=A_{2}$               &0.5                              & assumed     \\
$x_{1bol}$                  &0.480                            & assumed     \\
$x_{2bol}$                  &0.536                            & assumed     \\
$x_{1V}  $                  &0.506                            & assumed     \\
$x_{1R}  $                  &0.414                            & assumed     \\
$x_{2V}  $                  &0.726                            & assumed     \\
$x_{2R}  $                  &0.600                            & assumed     \\
$T_{1}   $                  &6750K                            & assumed     \\
q ($M_2/M_1$ )              &0.518                            & 0.003       \\
$\Omega_{in}=\Omega_{2}$    &2.8944                           & --          \\
$\Omega_{out}$              &2.5907                           & --          \\
$T_{2}$                     &4822K                            & 7K          \\
$i$                         &$89.^{\circ}86$                &$0.^{\circ}35$\\
$\Omega_{2}$                &3.590                            & 0.008       \\
$r_{1}(pole)$               &0.3229                           & 0.0008     \\
$r_{1}(side)$               &0.3317                           & 0.0009     \\
$r_{1}(back)$               &0.3402                           & 0.0010     \\
$r_{2}(pole)$               &0.3025                           & 0.0004     \\
$r_{2}(side)$               &0.3159                           & 0.0005     \\
$r_{2}(back)$               &0.3483                           & 0.0005      \\
$\Sigma{(O-C)^2}$           &0.0022                           &            \\
\hline\noalign{\smallskip}
 \end{tabular}
 \end{center}
 \end{table}

\begin{figure}[h!!!]
\centering
\includegraphics[width=9.0cm,angle=0]{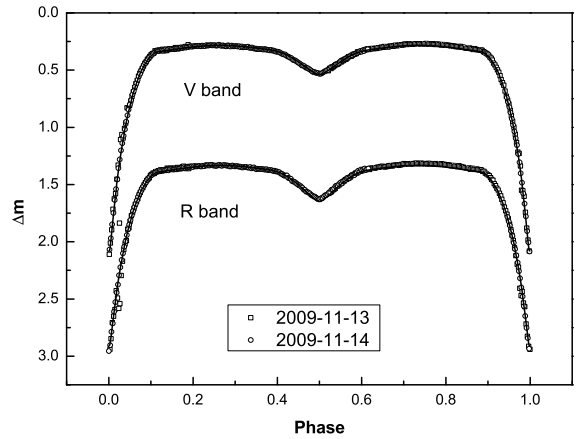}
\begin{minipage}[]{85mm}
\caption{Same as the top panel of Figure 1. But the solid curves
represent the theoretical light curves computed with the
parameters in Table 4. }
\end{minipage} \label{Fig4}
\end{figure}

\section{Discussion and Conclusions}
\label{sect:discussion}

In this paper, my photometric solution reveals that PS Per is a
semi-detached system. The Roche-geometry configuration that the
less massive and cool secondary component fills its inner Roche
lobe permits a dynamical mass transfer from the secondary to the
more massive primary star, suggesting a continuous period increase
just as in AI Cru (Zhao et al. 2010) and DD Mon (Qian et al.
2009). The orbital period of PS Per, however, does not show
continuous increase in this paper. This may be due to the low
quality of photographic and visual times, and the short span of
photoelectric and CCD times. In order to confirm the mass transfer
from the secondary to the primary star, long-term orbital timing
data are required.

According to the derived physical parameters listed in Table 4 and
the Harmanec's (1988) relation for masses and radii as functions
of spectral type, the following orbital parameters can be derived:
$M_1=1.31~M_{\odot}$, $R_1=1.39~R_{\odot}$, $L_1=3.62~L_{\odot}$,
$M_2=0.68~M_{\odot}$, $R_2=1.35~R_{\odot}$, $L_2=0.89~L_{\odot}$,
and $a=4.19~R_{\odot}$. In order to further verify the parameters,
Spectroscopic observations of the radial velocity curves of both
components are needed.

As showed in Figure 1, the second maxima of the light curves are a
little higher than the primary maxima. The weak O'Connell effect
maybe arises from a hot spot on the primary component as a result
of the impact of the gaseous stream from the cooler, less massive
secondary component. Such hot spot is often seen in other
semi-detached binary systems, such as CL Aur (Lee et al. 2010) and
KQ Gem (Zhang 2010). Considering the late spectral type and fast
rotation of the secondary star, The asymmetry of the light curves
can be also attributed to a cool spot on the secondary star caused
by magnetic activity. It is a reasonable trial that the magnetic
activity makes the light curves show more variability than the
impact of the gaseous stream does. So, in order to tell the two
mechanisms of magnetic activity and impact of the gaseous stream,
the investigation on long-term behaviour of the light curves is
also needed.

\normalem
\begin{acknowledgements}
I thank an anonymous referee for some useful suggestions. This
work is supported by Natural Science Foundation of Shanxi Normal
University (No. ZR09002).
\end{acknowledgements}

\label{lastpage}


\begin{thebibliography}{99}
\small \setlength{\itemindent}{-3mm} \setlength{\itemsep}{-0.5mm}
\setlength{\baselineskip}{4.5mm}
%% you can type \apj for ApJ, \aap for A&A, \apss for Ap&SS, etc. Please consult
%% the macro raa.cls. You can also find them in aasguide.tex (AASTeX for ApJ, AJ, PASP)
%% Please follow the formats of RAA's references list as demonstrated below:

\bibitem[{Agerer \& Hubscher} (2002)]{Agerer02}Agerer, F., \& H\"{u}bscher, J. 2002, \ibvs, No. 5296

\bibitem[{Agerer \& Hubscher} (2003)]{Agerer03}Agerer, F., \& H\"{u}bscher, J. 2003, \ibvs, No. 5484

\bibitem[{Brat}{et~al.} (2009)]{Brat09}Br\'{a}t, L., Trnka, J., \& Lehk\'{y},
M. 2009, OEJV, No. 107

\bibitem[{Diethelm} (2003)]{Diethelm03}Diethelm, R. 2003, \ibvs, No. 5438

\bibitem[{Diethelm} (2004)]{Diethelm04}Diethelm, R. 2004, \ibvs, No. 5543

\bibitem[{Diethelm} (2005)]{Diethelm05}Diethelm, R. 2005, \ibvs, No. 5653

\bibitem[{Diethelm} (2005)]{Diethelm05}H\"{u}bscher, J., Paschke, A., \& Walter,
F 2006, \ibvs, No. 5731

\bibitem[{Diethelm} (2010)]{Diethelm10}Diethelm, R. 2010, \ibvs, No. 5920

\bibitem[{Harmanec} (1988)]{Harmanec88}Harmanec, P. 1988, Bull. Astron. Inst. Czechoslovakia, 39, 329

\bibitem[{Kukarkin}{et~al.}(1968)]{Kukarkin68}Kukarkin, B. V., Efremov, Yu. N., Frolov, M. S., et
al. 1968, \ibvs, No. 311

\bibitem[{Kreiner}{et~al.}(2001)]{Kreiner01}Kreiner, J. M., Kim, C.-H., \& Nha, I.-S. 2001, An Atlas of $O -
C$ Diagrams of Eclipsing Binary Stars (Krak\'{o}w: Wydawnictwo
Naukowe Akad. Pedagogicznej)

\bibitem[{Kreiner}(2001)]{Kreiner04}Kreiner, J. M. 2004, Acta Astron., 54, 207


\bibitem[{Lee}{et~al.}(2010)]{Lee10}Lee, J. W., Kim, C.-H., Kim, D. H., Kim, S.-L., Lee, C.-U., \& Koch, R.
H. 2010, AJ, 139, 2669

\bibitem[{Qian}{et~al.}(2009)]{Qian09}Qian, S.-B., Zhu, L.-Y., Boonrucksar, S., Xiang, F.-Y., \& He,
J.-J. 2009, PASJ, 61, 333

\bibitem[{Safar \& Zejda} (2000)]{Safar00a}\v{S}af\'{a}\v{r}, J., \& Zejda M. 2000a, \ibvs, No. 4887

\bibitem[{Safar \& Zejda} (2000)]{Safar00b}\v{S}af\'{a}\v{r}, J., \& Zejda M. 2000b, \ibvs, No. 4888

\bibitem[{Hamme} (1993)]{Hamme93}van Hamme, W. 1993, AJ, 106, 2096

\bibitem[{Wilson \& Devinney} (1971)]{Wilson71}Wilson, R. E., \& Devinney E. J. 1971, ApJ, 166, 605

\bibitem[{Wilson} (1979)]{Wilson79}Wilson, R. E. 1979, ApJ, 234, 1054

\bibitem[{Wilson} (1990)]{Wilson90}Wilson, R. E. 1990, ApJ, 356, 613

\bibitem[{Zejda} (2002)]{Zejda02}Zejda, M. 2002, \ibvs, No. 5287

\bibitem[{Zejda} (2004)]{Zejda04}Zejda, M. 2004, \ibvs, No. 5583

\bibitem[{Zejda}{et~al.} (2006)]{Zejda06}Zejda, M., Mikul\'{a}\v{s}ek, Z., \& Wolf,
M. 2006, \ibvs, No. 5741

\bibitem[Zhang (2010)]{Zhang10}Zhang, L.-Y. 2010, PASP, 122, 309

\bibitem[{Zhao}{et~al.} (2010)]{Zhao10}Zhao, E.-G., Qian, S.-B., Fern\'{a}ndez Laj\'{u}s, E., von
Essen, C., \& Zhu, L.-Y. 2010, RAA(Research Astron. Astrophys.),
10, 438

\end{thebibliography}
\end{document}